\documentclass{article}

\usepackage{PRIMEarxiv}
\usepackage{booktabs}  
\usepackage[utf8]{inputenc} % allow utf-8 input
\usepackage[T1]{fontenc}    % use 8-bit T1 fonts
\usepackage{hyperref}       % hyperlinks
\usepackage{url}            % simple URL typesetting
\usepackage{amsfonts}       % blackboard math symbols
\usepackage{nicefrac}       % compact symbols for 1/2, etc.
\usepackage{microtype}      % microtypography
\usepackage{lipsum}
\usepackage{fancyhdr}       % header
\usepackage{graphicx}       % graphics
\graphicspath{{media/}}     % organize your images and other figures under media/ folder
\usepackage{pgfplots}
\usepackage{atbegshi}
\usepackage{caption}
\usepackage[multiple]{footmisc}
\usepackage{amsmath}
\usepackage{multirow}
\usepackage{amssymb}
\usepackage{subcaption}
\usepackage{xcolor}
\usepackage[normalem]{ulem}
\usepackage{comment}
\usepackage{algorithm}
\usepackage[noend]{algpseudocode}
\newcommand{\argmin}{\mathop{\mathrm{argmin}}\limits}
\usepackage{mathtools}
\let\norm\undefined % <-- "Undefine" \norm
\DeclarePairedDelimiter\norm{\lVert}{\rVert}
\usepackage[normalem]{ulem}
\title{Data-driven Trust Bootstrapping for Mobile Edge Computing-based Industrial IoT Services
}

\author{
Prabath Abeysekara\\
 Hitachi Construction Machinery,\\
 Brisbane, QLD 4076, Australia\\    
\texttt{email:prabathabeysekara@gmail.com} \\
  \AND
 Hai Dong \\
 School of Computing Technologies\\
  RMIT University, Melbourne, Australia\\
 \texttt{email:hai.dong@rmit.edu.au} \\
}

\begin{document}
\maketitle

\begin{abstract}
We propose a data-driven and context-aware approach to bootstrap trustworthiness of homogeneous Internet of Things (IoT) services in Mobile Edge Computing (MEC) based industrial IoT (IIoT) systems. The proposed approach addresses key limitations in adapting existing trust bootstrapping approaches into MEC-based IIoT systems. These key limitations include, the lack of opportunity for a service consumer to interact with a \textit{lesser-known} service over a prolonged period of time to get a robust measure of its trustworthiness, inability of service consumers to consistently interact with their peers to receive reliable recommendations of the trustworthiness of a lesser-known service as well as the impact of uneven context parameters in different MEC environments causing uneven trust environments for trust evaluation. In addition, the proposed approach also tackles the problem of data sparsity via enabling knowledge sharing among different MEC environments within a given MEC topology. To verify the effectiveness of the proposed approach, we carried out a comprehensive evaluation on two real-world datasets suitably adjusted to exhibit the context-dependent trust information accumulated in MEC environments within a given MEC topology. The experimental results affirmed the effectiveness of our approach and its suitability to bootstrap trustworthiness of services in MEC-based IIoT systems.
\end{abstract}

% keywords can be removed
\keywords{
Trust Bootstrapping  \and Mobile Edge Computing \and  Internet of Things Services \and  Distributed Machine Learning}

\section{Introduction}\label{sec:introduction}
Mobile Edge Computing (MEC)-based Industrial Internet of Things (IIoT) systems have gained significant attention in the recent past from academia and enterprises alike. Such a development has been motivated by the ability of MEC to tackle various challenges posed by the explosive growth of IoT devices on existing centralized IIoT systems. These challenges include tackling the ever-growing stress on the mobile networks caused by the high-volume IoT data, as well as facilitating delay-sensitive applications such as autonomous cars in Intelligent Transport System (ITS) settings. 
% To address the aforementioned challenges, MEC provides scalable and geographically distributed computing, storage as well as networking resources \cite{RN402} to host localized IIoT services at the edge of the network, sitting in close proximity to the services and their consumers \cite{RN407}.
To address the aforementioned challenges, MEC provides scalable and geographically distributed computing, storage, as well as networking resources \cite{RN402}\cite{10036140} to host geolocalized IIoT services at the edge of the network \cite{RN407}\cite{9583862}.

Despite the advantages, the system architecture of MEC and the challenges that come with it can cause these IIoT services hosted within different MEC environments to perform differently. This leads to service consumers requiring ways to assess these services and select those \textit{that best meet their requirements and expectations} before consuming them. Within the scope of this work, we refer to \textit{the ability of an IIoT service to meet the requirements and expectations of its consumers as its trustworthiness}. For instance, let us take an Intelligent Transport System (ITS) in an MEC-based IIoT eco-system \cite{xiong2019personalized}. This MEC-based ITS can provide traffic sensing services in the form of mobile crowdsensed services from vehicles with sensing capabilities as well as services provided by mobile traffic sensing Unmanned Aerial Vehicles (UAVs) and surveillance camera infrastructures \cite{xiong2019personalized}\cite{li2020novel}\cite{bai2024fedqtrustefficientdatadriventrust}\cite{10248268}. Interested consumers such as autonomous cars can consume these traffic sensing services to learn about congestions in close proximity as well as derive navigation decisions. In such an inherently complex environment where centralized authentication is often infeasible, the existence of malicious bodies within some MEC environments is inevitable\cite{RN403}\cite{9778269}. Ranging from botnets to compromised sensors, these malicious bodies can feed in fabricated information to the MEC-localized IIoT services causing undesirable effects to the service consumers consuming them \cite{RN379}\cite{9599374}. In addition, varying QoS characteristics \cite{RN404}\cite{8653299} of these services can also cause unhappy service consumers \cite{RN402}\cite{10036140}. As a result, the consumers of IoT services in MEC-based IIoT systems require strategies that help them better estimate the ability of a given service to cater to their requirements.

% In addition to the aforementioned systems challenges, the architecture of MEC-based IIoT systems as well as behavioural characteristics of the services (i.e. trustors) hosted in them and their consumers (i.e. trustees) challenge the adaptability of existing trust evaluation methods. We observe the following challenges that exist in such a setting.

However, the systems architecture of MEC-based IIoT systems, behavioural characteristics of the services (i.e. trustors) hosted in them and their consumers (i.e. trustees) challenge the adaptability of existing trust evaluation methods to achieve the aforementioned goal. We observe the following challenges that exist in such a setting.

\noindent\textit{\textbf{Challenge 1:} A trustee might not always have the opportunity to interact with a trustor for a prolonged period of time to get a reliable measure of \textit{direct trust}:} The trust between a trustor and a trustee engaged in a mutually beneficial relationship is most reliably determined when they interact with a prolonged period of time \cite{RN389}\cite{8523670}. This helps the trustees better understand the trustor and make more informed trust decisions while interacting with them. However, the dynamism of MEC-based IIoT systems prevents such prolonged relationships due to multiple factors. 
% For instance, services deployed in MEC-based IIoT environments could be short-lived as well as sporadically appear and disappear in an MEC environment due to the highly dynamic nature of their providers. As a representative example, 
For instance, an autonomous vehicle acting as a sensor data provider in an MEC-based ITS could enter and leave the coverage area of a given MEC environment within a short period of time \cite{RN406}. This can cause new short-lived sensor services to appear and disappear within a given MEC environment sporadically.  
%Such a phenomenon, coupled with the mobility of these dynamic services, 
Therefore, a mobility-enabled service consumer mobilizing past an MEC environment might not have enough knowledge gathered for a sufficient amount of time to accurately determine the trustworthiness (i.e. direct trust) of these lesser-known services using its direct experience with them. 

\noindent\textit{\textbf{Challenge 2:} A trustee might not always be able to directly communicate with its peers to evaluate the \textbf{indirect trust} of a trustor:} Most existing trust bootstrapping methods proposed for IoT systems, in general, rely on direct correspondence amongst a service consumer and its peers to determine the trustworthiness of a lesser-known service or its provider. Such approaches often attempts to evaluate the reputation of a service or its provider as observed by the other service consumers (i.e. indirect trust) \cite{RN394}\cite{RN392}\cite{8456378}\cite{10.1007/978-3-030-02925-8_13}. However, in an MEC setting, mobility-enabled service consumers (e.g. autonomous vehicles in an MEC-based ITS) who have prior experience interacting with a given service might not be available in abundance to communicate with or even exist in order to bootstrap an accurate \textit{priori trust} towards lesser-known services. Therefore, traditional trust bootstrapping methodologies that focus on evaluating indirect trust towards a given IIoT service by allowing service consumers to communicate directly with each other over a prolonged period of time can be deemed infeasible.

\noindent\textit{\textbf{Challenge 3:} Uneven contextual parameters in different MEC environments may demand the trustworthiness to be evaluated differently:} Trustworthiness of an IIoT service available within a given MEC environment may depend on multiple factors. These factors include functional properties and non-functional properties such as Quality of Service (QoS) characteristics of these services, which are well-known \cite{RN396}. In addition, there are other lesser-acknowledged factors such as operational characteristics, available computing and storage resources, channel conditions, etc, that tend to be different from one MEC environment to another. These conditions can influence the QoS characteristics of even the same type of services to be different among different MEC environments \cite{RN405}.  Such a behaviour, in turn, gives rise to different trust information distributions (i.e. \textit{non-identically and independent}, or in other words, \textit{non-IID} trust information distributions) \cite{9590330}. Consequently, the trustworthiness of even the same type of service is different across different MEC environments. Therefore, trust bootstrapping needs to be carried out in a \textit{data-driven} and \textit{context-aware} manner, adhering to the specific trust characteristics of each MEC environment.

\noindent\textit{\textbf{Challenge 4:} Split coverage area can cause context-aware trust information sparsity:} Most existing trust bootstrapping as well as evaluation strategies operate from centralized cloud-based infrastructures. Consequently, all trust information generated from the transactions between IIoT services and their consumers are accumulated in cloud-based centralized data centers. In contrast, each MEC environment accumulating the trust information from the transactions between MEC-local IIoT services and their consumers, only sees a split view of the world. While this allows establishing context-aware trust bootstrapping, the resulting trust information sparsity can hinder their ability to train reasonably accurate and generalisable prediction models to bootstrap the trustworthiness of a lesser-known IIoT service.

Distributed machine learning approaches based on \textit{edge-cloud collaboration} have emerged in the recent past as a useful paradigm for efficient predictive data analytics in MEC systems \cite{bai2024fedqtrustefficientdatadriventrust}\cite{10248268}\cite{9599374}. Such approaches promise significantly lower network stress on the core mobile network by collecting and processing data at the edge of the network while also utilizing cloud resources for a variety of tasks \cite{securefedlearning}\cite{RN323}. These tasks include knowledge aggregation and sharing amongst MEC environments, algorithm coordination, etc. \cite{RN323}. Therefore, to address the challenged elaborated previously, we propose \textit{a data-driven trust bootstrapping strategy for dynamic and lesser-known MEC-based IIoT services} using the distributed optimization paradigm atop \textit{edge-cloud collaboration}. 
% More specifically, the proposed approach
More specifically, we present the following contributions in this work.

% % \vspace{-1mm}
\begin{itemize}
    \item We formally model the problem of \textit{region-based trust bootstrapping} for MEC-based IIoT services as a distributed optimization problem in a way that it 
    \begin{itemize}
        \item allows using historical trust information gathered from the transactions among homogeneous IIoT services within a given MEC environment to determine the trustworthiness of a lesser-known service to counter the effects outlined by \textbf{challenge 1 and 2}.
        \item allows modelling trust characteristics of different MEC environments in a data-driven and context-aware manner as outlined in \textbf{challenge 3}.
        \item allows knowledge sharing within similar trust regions to counter the effects of context-aware trust information sparsity that can arise within MEC environments as elaborated in \textbf{challenge 4}. 
    \end{itemize}
    \item we also introduce a distributed and parallel algorithm to solve the aforementioned formulation collaboratively and in parallel using the Alternating Method of Multipliers (ADMM) framework by sharing knowledge among similar \textit{trust regions}. A \textit{trust region} refers to an MEC environment where \textit{a region-specific trust prediction model} can be established to determine a suitable \textit{priori trust} of a lesser-known service in response to trust queries from service consumers. This allows minimal data movement through the core networks of mobile network providers adhering to the goals of the MEC paradigm.
    % \item allows self-organization of the trust bootstrapping models in each MEC environment to facilitate efficient knowledge sharing amongst different MEC environments. This avoids having to set-up knowledge sharing topologies manually and (or) statically, which does not scale well in the context of MEC environments.
    \item Finally, we present the results of a comprehensive and exhaustive evaluation carried out in order to verify the ability of the proposed approach to tackle the challenges outlined. The aforementioned evaluation was carried out atop two real-world datasets curated suitably to demonstrate the characteristics of a MEC topology. In addition to that, we also evaluated the computational efficiency as well as scalability of the proposed approach to further assert its applicability within the outlined setting. 
\end{itemize}

% % \vspace{-1mm}

The rest of the paper is structured as follows: Section \ref{sec:related-work} reviews the prior research our work builds on. Section \ref{sec:problem-formulation} formally defines the problem setting we focused on, and conceptually models a mathematical framework to address the trust bootstrapping problem in MEC-based IoT services. Section \ref{sec:solution} details out the proposed solution and Section \ref{sec:evaluation} comprehensively documents the experiments carried out to evaluate the proposed solution. Section \ref{sec:conclusion} concludes our work and discusses possible future work.

\section{Related Work}\label{sec:related-work}

Existing literature introduces the problem of trust bootstrapping as the process of establishing a trust relationship between two entities in the form of a trustor and a trustee when there has been limited or no information available to reliably determine the trust between them \cite{RN389}\cite{8523670}. It remains a well-acknowledged and well-studied aspect particularly in relation to the cold-start problem associated with many existing trust evaluation strategies. Most existing approaches aim to tackle the cold-start problem arising as a result of little or no information available on an arbitrary trustor (irrespective of the application context) by directly talking to trustworthy neighbours of a given trustee \cite{RN386}\cite{9583862}. There is a clear \textit{paucity} in research that investigates the problem of trust bootstrapping in MEC-based services taking into account the inherent characteristics of such systems. Therefore, we review existing work related to trust bootstrapping in existing application contexts and identify their key limitations.

Thus far, trust bootstrapping in IoT systems has primarily been looked at in the current literature from a point-to point (e.g. Device-to-device, etc.) perspective. For instance, \cite{RN392} proposed a trust bootstrapping approach that takes into account influence from both trustors and trustees, which is then translated into a metric named trust propensity. In addition, \cite{RN394} and \cite{RN392} also propose approaches that can address the cold start problem in trust in the context of IoT systems, which involves evaluating the reputation of a device via the reputation information gathered directly from other devices. A key limitation of these approaches is that, the reputation of a service provider (i.e. another device or an application providing the desired service) is enquired by a given service consumer (i.e. IoT device seeking a service from a service provider) from other service consumers via direct communication. This, however, assumes that the topology involving the service providers and consumers is more-or-less static (e.g. the devices are stationary) or stays intact at all times. The aforementioned assumption does not hold true for highly dynamic MEC-based IoT systems, particularly in scenarios where \textit{service consumers do not enjoy the luxury of directly communicating with each other due to their mobility or inability to verify the trustworthiness of other devices}. \cite{RN393}, on the other hand, introduced an approach to bootstrap trust for D2D communication via reputation information gathered from a Profiling Server (PS) in an non-D2D manner. This PS, alongside the core network of a mobile network provider forms an extended network, which keeps track of reputation information of the device. However, to accomplish the proclaimed benefits, the aforementioned approach needs to accumulate trust information generated from each and every device within the PS sitting behind the core networks of mobile network providers. \textit{This helps little in terms of avoiding the network stress caused by devices deployed in large numbers}. Furthermore, \textit{none of the aforementioned approaches allows trust prediction to be done in a data-driven and context-aware manner}, which is another key limitation hindering their applicability in the context of trust bootstrapping within MEC-based IoT systems.

\section{Problem Formulation}\label{sec:problem-formulation}

Assume $M_{p} \in M$ to denote an arbitrary MEC environment with a service landscape as depicted in Fig. \ref{fig1}. We will, then, denote the different homogeneous service communities that exist in $M_{p}$ as $S_{p} = \{S_{p}^{q} | j \in \mathbb{N}_{>0}, 1 \leq q \leq n_{S_{p}}\}$. In this particular context, an $S_{p}^{q}$ represents a homogeneous, or in other words, functionally similar group of services that belong to a particular category (e.g. mapping, traffic information or parking services within an MEC environment that facilitates autonomous driving). 

\begin{figure}[t]
\centering
\includegraphics[width=\textwidth]{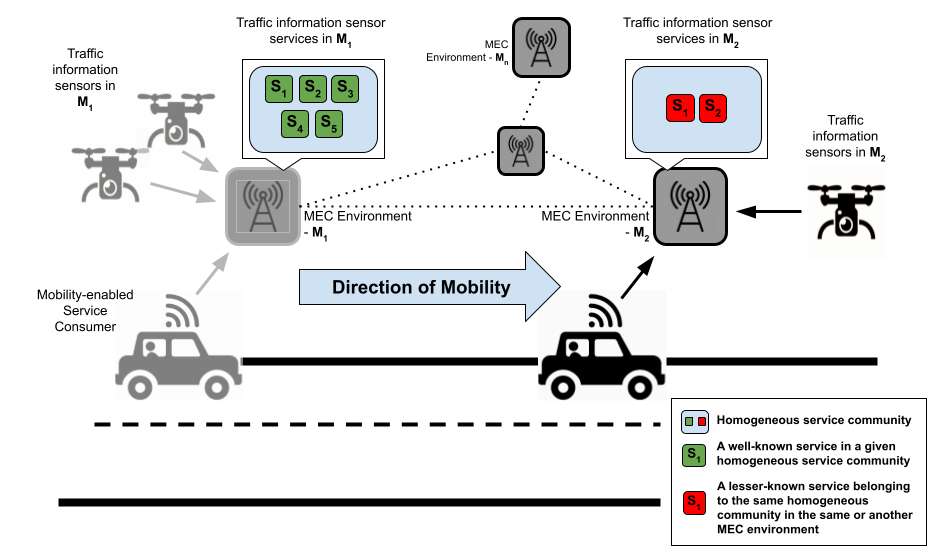}
\caption{A hypothetical illustration of the service landscape within a given MEC environment in an MEC-based IIoT system with respect to the autonomous vehicle use-case. 
} \label{fig1}
\end{figure}

Let us further extend the definition of $S_{p}^{q}$ as $S_{p}^{q} = \{s_{r} \cup {l}_{t} | r,t \in \mathbb{N}_{>0}\}$ to denote the set of functionally similar services belonging to it. In the aforementioned formulation, $s_{k}$ denotes a known service, or in other words, a service within an $S_{p}^{q}$ that carries sufficient historical trust information to determine its trustworthiness within a given MEC environment. $l_{t}$, on the other hand, denotes a lesser-known service within a MEC environment, or in other words, a service within $S_{p}^{q}$ that does not carry sufficient historical trust information to determine its trustworthiness. 

Given the trust information distribution corresponding to a given $S_{M_{p}}^{q}$ that constitutes all services within the aforementioned service community is represented by $\mathcal{P}_{S_{p}}^{q} = \{x, y\}_{i=n}^{i=1}$ where $x \in \mathbb{R}^{d}$, the problem of deriving a suitable \textit{priori trust} $BTrust_{l_{t}}$, or bootstrapping the trust of a lesser-known service $l_{t}$ could be formulated as

\begin{equation}\label{eq:bt-predictor}
   BTrust_{l_{t}} = f_{S_{p}}^{q}(x_{l_{t}}, w_{S_{p}}^{q})
\end{equation}

where $x_{l_{t}}$($\in \mathbb{R}^{d}$) denotes a comprehensive description of $l{t}$ representing the set of discriminative properties that defines trustworthiness of a given service belonging to the homogeneous service community $S_{p}^{q}$ in $M_{p}$, $w_{S_{p}}^{q}$ denotes the coefficients that define the extent to which each trust property in $x_{l_{t}}$ contributes to the trustworthiness of a service in $S_{p}^{q}$, and $f_{S_{p}}^{q}$ ($\in \mathcal{F}: \mathbb{R}^{d} \Rightarrow \mathbb{R}$) denotes a function (out of a family of functions represented by $\mathcal{F}$) that describes how the aforementioned properties and their coefficients are aggregated together producing a quantitative value that represents a priori trust for a lesser-known service $l_{t}$. Many existing research had used a linear combination of $x_{l_{t}}$ and $w_{S_{p}}^{q}$ to determine the trustworthiness of an IoT service, and therefore, we resort to $f_{S_{p}}^{q} = (w_{S_{p}}^{q})^{T} \cdot x_{l_{t}}$ in this work.

% To determine a suitable $w_{S_{p}}^{q}$ and $f_{S_{p}}^{q}$, we employ empirical risk minimization from machine learning theory, and define the following problem.
By applying basic machine learning theory, a $w_{S_{p}}^{q}$ that best matches a given trust information distribution $\mathcal{P}_{S_{p}}^{q}$ can be derived by minimizing the cumulative loss incurred by a suitable loss function $\ell$ such as Support Vector Machine (SVM), Least-squares, Linear or Logistic Regression) \cite{4635214}, as below. 
% \begin{equation}
%   w_{S_{p}}^{q} =  \underset{w \in \mathbb{R}^{d}}{\text{minimize}} & \frac{1}{N}\sum\limits_{j=1}^{N}f_{i}(w; \xi_m^{j})
% \end{equation}
% % \vspace{-1mm}
\begin{equation}\label{eq:loss-function}
\begin{array}{ccclcl}
w_{S_{p}}^{q} = \underset{w \in R^{d}}{\text{minimize}} & \displaystyle\sum\limits_{i=1}^{n} \ell(\mathcal{P}_{S_{p}}^{q}; w) = \underset{w \in R^{d}}{\text{minimize}} & L(w)
\end{array}
\end{equation} 

Within a given MEC topology, we can derive a set of predictors (as denoted by \eqref{eq:bt-predictor}) to bootstrap the trustworthiness of a lesser-known service by solving the problem \eqref{eq:loss-function} for a given homogeneous service community $S_{p}^{q}$ in parallel against each MEC environment and atop MEC-local trust information distributions $\mathcal{P}_{S_{p}}^{q}$ within the underlying MEC topology $M$. This will result in a series of \textit{context-aware} trust bootstrap models denoted as $W_{S_{p}} = \{w_{S_{p}}^{1}, w_{S_{p}}^{2}, \dots, w_{S_{p}}^{m}\}$. However, as described in \textbf{challenge 4} of Section \ref{sec:introduction}, the sparsity of trust information can hamper the ability of a trust bootstrap model to reliably determine the trustworthiness of a lesser-known service within some MEC environments. To alleviate the aforementioned challenge, by allowing MEC environments to collaborate, we modify the problem \eqref{eq:loss-function}, as below.

% \begin{equation}\label{eq:reg-loss-function}
% \begin{array}{c}
% w_{S_{p}}^{q} = \underset{w \in R^{d}}{\text{minimize}} L(w) + \gamma G(w, \{w_{i}\}_{w \neq w_{i}}) \\
% \text{where}\\
% G = \norm{w - \frac{1}{|N(i)|}\sum\limits_{i \in N(M_{p})} w_{i}}_{2}
% \end{array}
% \end{equation}

\begin{equation}\label{eq:reg-loss-function}
\begin{array}{c}
w_{S_{p}}^{q} = \underset{w \in R^{d}}{\text{minimize}} L(w) + \gamma G(w, \{w_{i}\}_{w \neq w_{i}}) \\
\text{where}\\
G = \sum\limits_{M_{q} \in N(M_{p})} \frac{n_q}{d(M_{p}, M_{q})}\norm{w - w_{q}}_{2}
\end{array}
\end{equation}

where $G$ infuses the knowledge (i.e. model parameters) extracted from the neighbours. $G$ encourages the parameters $w_{S_{p}}^{q}$ of a trust bootstrap model within an MEC environment to be selected from the knowledge acquired from its neighbours ($M_{q} \in N(M_{p})$ either by adopting their entire model or an aggregated form of (e.g. mean) the model parameters of the neighbours, under different circumstances. Meanwhile, $\gamma$ scales $L(w)$ with respect to $G$. In other words, it helps determine if the solution $w$ should be more closer to what is derived atop the MEC-local dataset or the knowledge shared by the neighbors. Furthermore, $n_q$ represents the number of training samples in $M_{q}$ and $d(M_{p}, M_{q})$ represents a distance function, which measures the \textit{physical} distance between the MEC environments $M_{p}$ and $M_{q}$. Here, weighting the knowledge $w_{q}$ shared by $M_{q}$ by a factor of $n_q$, allows reducing the impact of \textit{knowledge sharing} MEC environments with sparse trust information attempting to share sub-optimal or potentially overfitted knowledge with \textit{knowledge seeking} MEC environments. $d(M_{p}, M_{q})$, on the other hand, allows giving more prominence to MEC environments that are in close proximity thereby sharing similar trust information. We hypothesize that service consumers are more likely to mobilize amongst nearby MEC environments \cite{RN323}. Thus, giving more prominence to the knowledge shared by such MEC environments can be deemed more relevant in the aforementioned setting.

However, $G$ spoils the parallelism enjoyed by \eqref{eq:loss-function} as it now depends on the model parameters of $M_p$'s neighbours, which need to be determined at the same time or before that of $M_p$. Therefore, we look to aggregate all sub-problems (denoted by \eqref{eq:reg-loss-function}) that are to be solved by each MEC environment together, as below and attempt to derive a parallelizable solution.
% % \vspace{-1mm}
\begin{equation}\label{eq:loss-function1}
\begin{array}{ll}
[w_{S_{1}}^{q}, w_{S_{2}}^{q}, \dots, w_{S_{m}}^{q}] =\\ 
\hfill \underset{w_{1}, w_{2}, \dots, w_{m} \in R^{d}}{\text{minimize}} \bigg(\sum\limits_{i = 1}^{m} L(w_{i}) + \sum\limits_{i = 1}^{m}\gamma_{i} G_{i}(w_{i}, \{w_{j}\}_{w_{i} \neq w_{j}}) \bigg)
\end{array}
\end{equation}

\section{Solution}\label{sec:solution}

This section provides a comprehensive overview of the proposed solution and the theoretical foundation upon which it is developed.

\subsection{Technical Preliminary}

For completeness, we provide a brief systematic exposition on the ADMM framework, below.
% the Network Lasso problem, 
% ADMM and Hierarchical Clustering, below that the proposed approach relies on.

\subsubsection{Alternating Direction Method of Multipliers (ADMM)}
ADMM allows a convex optimization problem of the shape \eqref{eq:admm} to be decomposed into multiple sub-problems and solve them in a coordinated manner (via passing only the model parameters among sub-problems, instead of raw data) to derive a global solution. The aforementioned property of ADMM, therefore, makes it a good fit for an inherently distributed MEC topology where it is preferable to isolate the processing of an MEC-local dataset (i.e. solving a single sub-problem) in a given MEC environment closer to where the data was originated, yet allowing the communication (i.e. to share knowledge) among other MEC environments (i.e. connected neighbours) via passing smaller messages (i.e. model parameters). ADMM intends to take on the problems of type,

\begin{equation}\label{eq:admm}
\begin{aligned}
& {\text{minimize}}
& & \displaystyle f(w) + g(z)  \text{s.t.}
& & \displaystyle Aw + Bz = c
\end{aligned}
\end{equation}
\noindent where $\left. w \in \mathbb{R}^{n}, z \in \mathbb{R}^{m}, A \in \mathbb{R}^{p\times n}, B \in \mathbb{R}^{p\times m} \right.$. It is  assumed that the functions denoted by $\left. f(w) \right.$ and $\left. g(z) \right.$ are convex and defined as $\left. f: \mathbb{R}^{n} \right.$ and $\left. g: \mathbb{R}^{m} \right.$ \cite{RN211}. In most convex optimization problems where ADMM is applied, $\left. f(w) \right.$ corresponds to a loss function whereas $\left. g(z) \right.$ corresponds to a regularization  term that helps better generalize the solution of the optimization problem being solved.

To solve the constrained optimization problem \eqref{eq:admm} as an unconstrained problem, the augmented Lagrangian associated with it $\left. L_{p}(w,z,\mu) \right.$ is obtained similar to \cite{RN192}. By applying dual-ascent iteratively, ADMM minimizes the augmented Lagrangian $\left. L_{p}(w,z,\mu) \right.$ with the following steps.

\begin{subequations}\label{eq:admm-subs}
  \begin{align}
    w^{k+1} & =\argmin_{w\in\mathbb{R}^{n}}L_{p}(w,z^{k},\mu_{k}) \label{current_rel1a} \\
    z^{k+1} & =\argmin_{z\in\mathbb{R}^{m}}L_{p}(w^{k+1},z,\mu_{k}) \label{current_rel1b} \\
    \mu^{k+1} & =\mu^{k} + \rho\nabla\mu L_{p}(w^{k+1},z^{k+1},\mu)
    \label{current_rel1c}
  \end{align}
\end{subequations}
\noindent where $\left. k \right.$ represents the iteration number. When $\left. f(w) \right.$ and $\left. g(z) \right.$ are separable into multiple sub-problems, each solved over a partition of their respective datasets, the aforesaid iterations can be carried out to solve each sub-problem independently in parallel. NL framework utilizes this feature to solve optimization problems on potentially large graphs.

\subsection{Our Solution}

In this subsection, we present the proposed parallel solution to derive a data-driven, context-aware and self-organizing trust bootstrapping model for MEC-based IoT systems.

\begin{figure*}[t]
\centering
  \includegraphics[width=\linewidth,keepaspectratio]{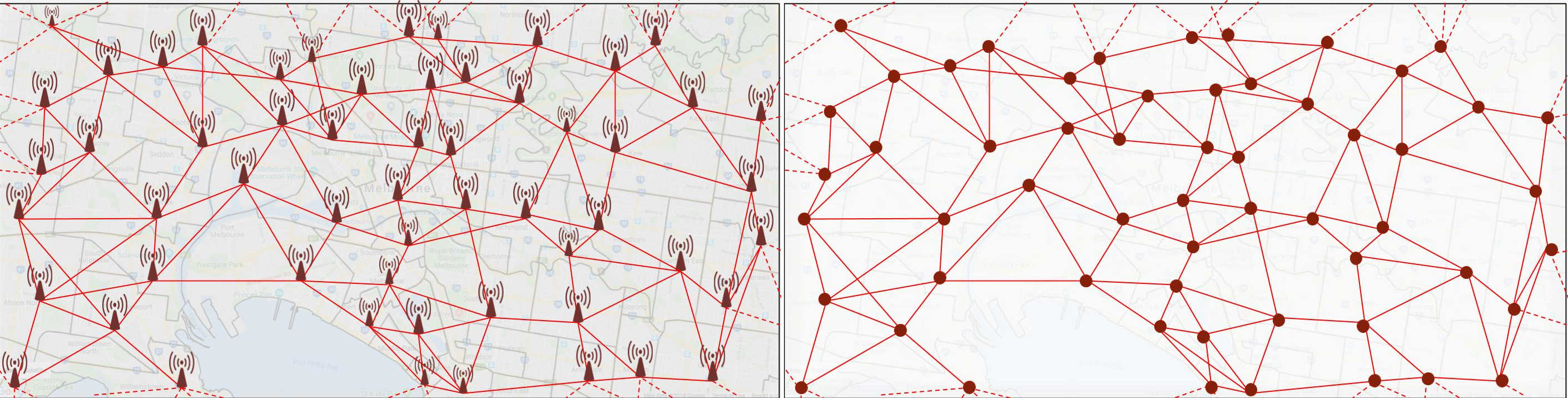}
  \caption{A hypothetical deployment of MEC environment, which shows how the neighbouring MEC environments are linked based on proximity forming a partial mesh network.}\label{fig2}
\end{figure*}

In a typical MEC topology, individual MEC environments tend to operate independently from others within their own network boundaries \cite{RN398}. This can hinder their ability to share knowledge with each other. In addition, although direct communication amongst the MEC environment for knowledge sharing is possible \cite{RN381}, complexities in inter-MEC network communication coupled with lack of interoperability standards encouraged us to utilize the centralized cloud to facilitate knowledge sharing. Even though the MEC paradigm attempts to overcome scalability challenges posed by centralized cloud-based infrastructure in the face of high-volume IoT data, \textit{edge-cloud collaboration} has attracted much attention in order to simplify communication among MEC environments \cite{RN188}. In such a setting, each MEC environment can be \textit{logically} connected to multiple MEC environments via the centralized cloud for collaborative training of trust bootstrapping models. The trust bootstrapping problem in MEC-based IoT systems formulated in Section \ref{sec:problem-formulation} was then modelled over the graph resulting from this topology (see Fig. \ref{fig2}), and Alternating Direction Method of Multipliers (ADMM) was applied to derive a parallel solution to train a distributed trust prediction model giving rise to Algorithm \ref{alg:solution}. ADMM allows a suitable convex optimization problem to be decomposed into multiple sub-problems and solve them in a coordinated manner (via passing only the model parameters among sub-problems, instead of raw data) to derive a global solution \cite{RN211}. This makes it a good fit for an inherently distributed MEC topology where it is preferable to isolate the processing of an MEC-local trust bootstrapping dataset (i.e. solving a single sub-problem) closer to where the data originated, also allowing the communication (i.e. to share knowledge) among other MEC environments (i.e. connected neighbours) via passing smaller messages (i.e. model parameters).

\begin{algorithm}[hbt!]
\caption{Data-driven, context-aware and self-organizing trust bootstrapping for MEC-based IoT services}\label{alg:solution}
\begin{algorithmic}[1]
    \State \textbf{inputs:} $M$-MEC environments, $e_{p}, e_{d}$-Appropriate thresholds for primal and dual residuals, $\rho$-Penalty parameter controlling constraint violations, $\gamma_{ini}$-Initial value of the parameter enforcing knowledge sharing, $\gamma_{inc}$-Factor by which $\gamma$ is incremented, $\gamma_{th}$-Stopping criteria for incrementing $\gamma$, $T$-Maximum number of iterations ADMM runs for.
    \vskip 4pt
    
    \Procedure{BOOTSTRAP}{$M, E, \gamma_{ini}, \gamma_{th}, \gamma_{inc}, \rho, T, e_{p}, e_{d}$}
        \State Initialize links among MEC environments for knowledge sharing - $E$
        \State Initialize all parameters - $W, Z, U \gets 0$
        % \comment\text{Initializes links among MEC environments for knowledge sharing} 
        \State $\gamma \gets \gamma_{init}$
    
        \ForAll{$m \in M$}\Comment\text{Loop over MECs in Cloud layer}
            \State Send initial $z_{ij}$, $z_{ji}$ and $u_{ij}$ to $m$ 
        \EndFor
        \vskip 4pt

        \While{$\gamma \leq \gamma_{thresh}$}
            \State $W, Z, U \gets\textsc{ADMM}(\gamma, \rho, T, e_{p}, e_{d}, W, Z, U)$
            \State $\gamma \gets \gamma * \gamma_{inc}$
        \EndWhile
        \State \textbf{return} $W$
    \EndProcedure
    \vskip 4pt
    
    \Procedure{ADMM}{$\gamma, \rho, T, e_{p}, e_{d}, W, Z, U$}
        \State Initialize primal and dual residuals $\left.res_{p}^{k} \mathbin{\gets} 0\right.$ and $\left.res_{d}^{k} \mathbin{\gets} 0\right.$
        \While{$\norm{res_{p}^{k}}_{2} \geq e_{p}; \norm{res_{d}^{k}}_{2} \geq e_{d}$}
          \ForAll{$m \in M$}\Comment\text{Distributed loop over MECs}
            % \State\begin{alignat*}{2}
            
            \State $w_{i}^{k+1} \gets\textsc{W-Update}(z_{ij}^{k}, u_{ij}^{k})$
            % \end{alignat*}
            % \State Send $\left. w_{i}^{k+1}\right.$ to cloud layer
          \EndFor
          \ForAll{$e \in E$}\Comment\text{Loop over the links among, MECs in Cloud}
            % \State\begin{alignat*}{}
            \State $z_{ij}^{k+1}, z_{ji}^{k+1} \gets\textsc{Z-Update}(w_{i}^{k+1}, u_{ij}^{k})$
            \State $u_{ij}^{k+1} \gets\textsc{U-Update}(w_{i}^{k+1}, z_{ij}^{k+1})$
          \EndFor
          \State Compute $\left.\norm{res_{p}^{k}}_{2}\right.$ and $\left.\norm{res_{d}^{k}}_{2}\right.$
        \EndWhile\label{euclidendwhile}
        \State \textbf{return} $W, Z, U$
    \EndProcedure
    \vskip 4pt

    \Procedure{W-Update}{$z_{ij}^{k}, u_{ij}^{k}$}
        \State $w_{i}^{k+1} \gets\argmin_{w_{i}} \Big(f_{i}(w_{i}) + \sum\limits_{j \in N(i)} \frac{\rho}{2}\norm{w_{i} - z_{ij}^{k} + u_{ij}^{k}}_{2}^{2}\Big)$
        \State \textbf{return} $w_{i}^{k+1}$ to the cloud layer
    \EndProcedure
    \vskip 4pt
    
    \Procedure{Z-Update}{$w_{i}^{k+1}, u_{ij}^{k}$}
        \State $z_{ij}^{k+1}, z_{ji}^{k+1} \gets\argmin_{z_{ij},z_{ji}}\Big(\frac{\gamma_{ij}.n_{j}}{d_{ij}}G(z_{ij},z_{ji}) + \frac{\rho}{2}(\norm{w_{i}^{k+1} - z_{ij} + u_{ij}^{k}}_{2}^{2} + \norm{w_{i}^{k+1} - z_{ji} + u_{ji}^{k}}_{2}^{2})\Big)$
        \State \textbf{return} $z_{ij}^{k+1}, z_{ji}^{k+1}$
    \EndProcedure
    
    \Procedure{U-Update}{$(w_{i}^{k+1}, z_{ij}^{k+1}$}
        \State $u_{ij}^{k+1} \& \gets u_{ij}^{k} + (w_{i}^{k+1} - z_{ij}^{k+1})$
        \State \textbf{return} $u_{ij}^{k+1}$
    \EndProcedure
\end{algorithmic}
\end{algorithm}

Algorithm \ref{alg:solution} runs in multiple key steps in harmony with the cloud and MEC layers. First, the model parameters of trust bootstrapping models trained by each MEC environment are initialized by a Global Model Coordinator (GMC) running in the cloud layer (see lines [4-5]). After that the ADMM procedure runs its three key steps 
%denoted by problem \eqref{eq:admm-subs} 
alternatingly between the cloud and MEC layers, as below.
 
\noindent\textbf{$w_{i}$-update}: Separable across each local MEC environment, $w_{i}$-update is solved iteratively in parallel atop MEC-local trust information. Utilizing the $z_{ij}$- and $u_{ij}$-updates from the previous iterations shared by the GMC during the initialization phase, each local MEC layer then independently trains its own local trust bootstrapping model (see line 18). Once done, all MEC environments share their resulting model parameters $w_{i}$ through the GMC (see line 19).

\noindent\textbf{$z_{ij}$-, $z_{ji}$- and $u_{ij}$-updates:} In contrast to $w_{i}$-update, $z_{ij}$-, $z_{ji}$- and $u_{ij}$-updates are carried out by the cloud layer. Out of the aforementioned steps, $z_{ij}$-, $z_{ji}$- perform knowledge sharing by forcing the model parameters of the trust bootstrapping model trained by a given MEC environment to be similar to the mean of the cluster it belongs to (see lines 13, [20-22]), while $u_{ij}$-updates concerns with updating dual variables used by the ADMM framework (see lines 14, [23-25]).

\noindent\textbf{Output:}  The output produced by the ADMM procedure (see line 16) after the aforementioned framework converges (or reaches an early-stopping, which often is the case as ADMM tends to slow-down as it reaches the optimum \cite{RN211}), consists of the model parameters of each individual MEC environment corresponding to their trust bootstrapping models.

We used a soft-margin SVM \cite{RN219} as the loss function (i.e. $f_{i}$) to be minimized in each $w_{i}$-update carried out by individual MEC environments, above (see line 17). SVM has already been widely used and shown to work well in prior trust research for modelling trustworthiness of IoT services \cite{RN323}. This background provided us with a rational basis to adopt SVM as the problem to be solved as part of each sub-task running in the local MEC layers (i.e. loss function to be minimized) of the reference implementation. In that, each local MEC environment trains its own SVM-based binary classifier to derive a priori trust for a given lesser-known IoT service. Each trained classifier classifies an input as either ``benign" or ``harmful" (denoted by ``1" and ``-1" respectively) indicating whether the lesser-known IoT service in question is trustworthy or not.  

% \vspace{-4mm}
\section{Evaluation}\label{sec:evaluation}
This section is organized as follows. Section \ref{subsec:experiments} details out the experiments designed to evaluate the proclaimed capabilities of the proposed approach and the Key Performance Indicators (KPIs) used to measure them. Section \ref{subsec:datasets} provides detailed descriptions of the datasets used for the evaluation, and how they have been carefully curated to test the strengths of the proposed approach. Meanwhile, Section \ref{subsec:models-compared} introduces all the baselines models compared whereas Section \ref{subsec:results} discusses the results obtained against the experiments described in Section \ref{subsec:experiments}.

% \vspace{-1mm}
% \noindent\textbf{Experiments:} 
\subsection{Experiments}\label{subsec:experiments}

% % \vspace{-1mm}
We conducted a series of experiments to evaluate the suitability of the proposed approach to address the challenges outlined in Section \ref{sec:introduction}. These experiments were grouped into three main categories, as below.

% % \vspace{-1mm}
\begin{itemize}
\item \textit{Effectiveness of data-driven and context-aware trust bootstrapping of IoT services within an MEC environment:} To evaluate this, the performance of the proposed approach was compared against a global trust bootstrapping model resembling a scenario where all service consumers share their trust information with a centralized global server for deriving the trustworthiness of a lesser-known service forming a single \textit{context environment} for trust bootstrapping. Accuracy was used as the primary key performance indicator (KPI) to compare the performance of each binary SVM classifier trained during the experiments.

\item \textit{Effectiveness of knowledge sharing:} To evaluate this, the proposed approach was compared against the non-collaborative baseline models outlined in Section \ref{subsec:models-compared}. Accuracy was again used as the primary KPI to compare the performance among the evaluated models.

\item \textit{Communication efficiency:} Although the proposed approach addresses key challenges impacting trust bootstrapping in MEC environments, it is also imperative that we assess its alignment with the goals of MEC. To evaluate this, we also measured the number of \textit{rounds of communication}, which is \textit{an indicative measure of the network stress on the core mobile networks}, needed during the end-to-end process that includes trust information accumulation and prediction model training between the centralized cloud and distributed MEC layers.

The communication efficiency was only compared amongst the proposed approach, GLB-TBM, Global TT-SVD and Global Wahab et. al's. This is because none of the other models required the data to be transmitted out of the network boundaries of the MEC environments within which the data was accumulated and model training took place. Therefore, there was no communication across the core networks of mobile network providers, and thus, no network stress on them.

\item \textit{Computational efficiency:} We primarily considered \textit{total running time-to-maximum accuracy} as the primary KPI of computational efficiency. To that end, we have compared the \textit{total wall-clock time} taken by the proposed approach as well as the other state-of-the-art trust bootstrapping models to achieve the maximum observed accuracy. For simplicity, \textit{we assumed that the communication overhead between MEC and cloud layers is negligible}. Although, in reality, a communication overhead has a significant effect on the overall computational efficiency of the proposed approach, we believe the aforementioned assumption provides a fair-ground to compare its performance against the other baseline and state-of-the-art non-distributed models. We provide implications of this approach in Section V.D. It is imperative to note that this experiment only aims to measure the time taken by each compared approach to train their respective models. In other words, we leave out the time taken for communicating intermediate outputs between the MEC and Cloud layers to allow a fair evaluation amongst the compared models on the computational efficiency.

\item \textit{Scalability:} These experiments were designed to evaluate the scalability of our approach with respect to the \textit{growth} of multiple aspects outlined below.
\begin{enumerate}
    \item \textbf{IIoT services and consumers:} From a proposed solution’s perspective, the growth of IIoT service and consumers directly translates to a growth in trust information generated by the interactions amongst them. Therefore, to assess \textit{the ability of our proposed solution to withstand growing IIoT services and consumers}, we evaluated its performance against MEC-local datasets of which the sizes were increased by 25\%, 50\%, 75\% and 100\%. The total running time and number of communication iterations until convergence of the proposed solution were used as the Key Performance Indicators (KPIs) to evaluate the performance of this aspect.
    \item \textbf{Number of MEC environments in the MEC topology:} To assess \textit{the ability to scale well to growing topology sizes}, we monitored the average prediction accuracy across all distributed trust prediction models in a given MEC topology as well as the average number of communication rounds required till convergence when the number of MEC environments in the underlying MEC topology is gradually increased. The other non-distributed state-of-the-art models were left out from this experiment as they use only a single global model that does not scale across a given MEC topology.
   %\item Number of connected neighboring MEC environments for knowledge sharing: 
\end{enumerate}
\end{itemize}

To reduce bias, the results presented have been either taken as the average of multiple rounds of experiments, or via cross-validation where appropriate.

\noindent\textbf{Experiment Set-up:} Extending the problem setting described in Section \ref{sec:problem-formulation} (see Fig. \ref{fig1} for an illustration), we designed a simulated experimental set-up scenario where there is a hypothetical MEC topology with 100 MEC environments across 100 suburbs in the Melbourne City Council area. We marked every MEC environment pertaining to a particular suburb as a node in a graph laid on top of a map of Melbourne City Council (see Fig. \ref{fig2}). Each MEC environment, then, trains a trust bootstrapping model. The setup primarily consisted of an application written in Python. This application was used to orchestrate and train a distributed trust prediction model using the proposed machine learning architecture. 

\subsection{Datasets}\label{subsec:datasets}

We used two public IoT datasets for our experiments. A comprehensive overview of the structure of these datasets is given below .

% % \vspace{-1mm}
\begin{itemize}
    \item \textbf{UNSW-NB15\footnote{https://www.unsw.adfa.edu.au/unsw-canberra-cyber/cybersecurity/ADFA-NB15-Datasets/}:} This dataset consists of transaction data collected from 43 \textit{pseudo sensors} (i.e. simulated sensors tagged with unique source IP addresses) in a simulated intrusion detection system (IDS). Each record in the dataset contains 49 numerical and categorical features, i.e. $\in \mathbb{R}^{49}$ and corresponds to a transaction indicating either a benign behaviour and or one of nine types of attack scenarios. We labelled each sample as \textit{benign} or \textit{harmful} based on whether they correspond to a benign or attack scenario. In addition, the categorial variables were converted into numerical variables using \textit{one-hot encoding} strategy \cite{cerda2018similarity}. This resulted in a dataset of the dimensionality $\mathbb{R}^{191}$.
    \item \noindent\textbf{N-BaIoT\footnote{https://archive.ics.uci.edu/ml/datasets/detection\_of\_IoT\_botnet\_ attacks\_N\_BaIoT}:} This dataset consists of network traffic data collected from 9 smart devices previously used to detect Mirai and BASHLITE attacks within an IoT setting. Under each family of attacks, there were multiple individual attack types of which the records ($\in \mathbb{R}^{115}$) were consolidated under the label \textit{harmful}. In addition, the records related to legitimate network traffic ($\in \mathbb{R}^{115}$) were classified under the label \textit{benign}.
\end{itemize}

Both the aforementioned datasets were scaling using sklearn MinMaxScaler \footnote{https://scikit-learn.org/stable/modules/generated/\\sklearn.preprocessing.MinMaxScaler.html} before the learning models were used. This was done in order to reduce the impact of features carrying values of higher magnitude dominating the training process.

\noindent\textbf{Dataset Preparation:}
% These datasets were first normalized and then divided into 100 randomly-sized (n$\left. \in \right.$[1000, 20,000]) smaller datasets. Random noise was also added to these datasets via flipping the labels of randomly picked samples to mimic a Non-IID dataset. The resulting datasets (with a training-to-test split ratio of 70:30) were used to train the distributed prediction models for each simulated MEC-environment. 
To simulate a justifiable environment for trust bootstrapping as per the setting described in Section \ref{sec:introduction}, we made a key assumption that the \textit{sensors} (i.e. smart devices) of the two trust datasets, can be directly represented as \textit{sensor services}. We consider this assumption to be pragmatic given the emerging paradigms such as sensor-as-a-service, where each sensor can be exposed as a service \cite{RN400}. We randomly split the dataset collected from each device into randomly-sized (n$\left. \in \right.$[1000, 20,000]) smaller datasets, each representing the trust information accumulated from an IoT service. Meanwhile, to form homogeneous communities of services, we added the same amounts of noise to groups of splits (i.e. data splits with different amount of noise correspond to different communities of homogeneous services). Lesser-known services were formed by creating considerably smaller data splits (n$\left. \in \right.$[100, 500]) compared to that of other simulated services.

% \textbf{(needs to explain in detail how to form a context-aware trust bootstrapping environment and simulate the lesser-known services)}

% \vspace{-4mm}
\subsection{Models Compared}\label{subsec:models-compared}

% % \vspace{-1mm}
This section provides an overview of the properties of each baseline model used for the evaluation. To the best of our observations, we did not come across any suitable trust bootstrapping approaches in existing literature that aim to tackle the same problems outlined in Section \ref{sec:introduction} in the context of MEC-based IoT systems. Therefore, we resorted to two models representing the abundantly available trust bootstrapping approaches previously proposed for centralized or D2D systems, which also rely on machine learning techniques.

\noindent\textbf{Wahab et. al’s} \cite{RN385}: A family of decision trees were trained corresponding to a simulated set of users taking part in the proposed trust bootstrapping strategy. Each decision tree was trained using $k$-fold ($k=10$) cross validation and uses the GINI algorithm to determine the best split for each node.

% \noindent\textbf{Aljazzaf et. al’s} \cite{RN389}:
% \vspace{1mm}

\noindent\textbf{TT-SVD} \cite{RN392}: 
% This approach takes a bidirectional approach to determine the trustworthiness of an entity with the basis that a trustor and trustee can always communicate with each other. However, in our setting the communication between the trustor (i.e. IoT service) and trustee (i.e. service consumer) is unidirectional from the latter to the former. 
This approach takes a bidirectional approach to determine the trustworthiness of an entity with the basis that in an event where a user (i.e. service consumer, in our context) has to determine the trustworthiness of a lesser-known item (i.e. IoT service), the bootstrapped trustworthiness is influenced by the other users (i.e. other service consumers) the user in question trusts and (or) is a trustee to. However, given the dynamism or MEC-based IoT systems described in Section \ref{sec:introduction}, and the difficulties in establishing direct trustworthy communications among other users, we relaxed the conditions that enforce the aforementioned aspects by setting $E_{u} = \{\}$, $T = [0]_{m \times m}$ and $E_{v}^{+} = \{\}$. In addition, we also set $\alpha = 0.1$ and $\mu = 5$ as per the authors' recommendation. We compared two variants of this approach in our problem setting, in the form of \textbf{Global TT-SVD} and \textbf{MEC-local TT-SVD} resembling two environments running TT-SVD in a centralized cloud as well as non-collaborative MEC setting, respectively.

% Therefore, we made an adjustment to the TT-SVD model to negate the impact of the TrusteeSVD component on the final recommendation it produces by setting $\beta = 0$ as below.

% % \vspace{-1mm}
% \begin{equation}
%   \begin{array}{ll}
%     r_{u, j} &= \beta (TrusteeSVD) + (1 - \beta) (TrustorSVD) \\
%              &= (TrustorSVD) 
%   \end{array}
% \end{equation}

% \textbf{(only one existing model is not enough. Also needs to provide justification why select those models)}
% % \vspace{-1mm}

In addition, we also compared our work against the following two variants of our proposed approach.

\begin{itemize}
\item \textbf{GLB-TBM:} A global trust bootstrapping model resembling a scenario where all service consumers share their trust information with a centralized global server for deriving the trustworthiness of a lesser-known service.
\item \textbf{LO-TBM:} A family of non-collaborative trust bootstrapping models resembling a scenario where trust information collected from service consumers are accumulated only within the MEC environment where they are operating from, and it is not shared outside the aforementioned MEC environment for knowledge sharing purposes.\end{itemize}

% \vspace{-5mm}
\subsection{Results and Discussion}\label{subsec:results}
This section provides comprehensive details on the results observed during the experiments in Section \ref{subsec:experiments} and their interpretations.

\subsubsection{Effectiveness of data-driven and context-aware trust bootstrapping}
The results of our experiments showed that that the context-aware binary SVM classifiers trained by the proposed approach consistently outperformed all the approaches \textit{made to} promote context-awareness across MEC environments (see TABLE \ref{tab:avg-pred-acc-over-num-nodes}).

We explain this behaviour by taking into consideration multiple contrasting aspects in the inner workings of the proposed approach and the other trust bootstrapping strategies evaluated. For instance, TT-SVD is based on Matrix Factorization. Even though Matrix Factorization is able to efficiently deal with the sparse user-rating data, they are not applicable to standard prediction data (e.g. a real valued feature vector in $\mathbb{R}^n$) \cite{RN401}. In other words, they rely on a set of latent features that are uncovered by factoring a dataset, which act as a profile for a given user (i.e. a service consumer, in our problem context). In a scenario where an IoT service is new, and there have not been many service consumers that had interacted with it previously, TT-SVD can find it unable to perform satisfactorily. On the other hand, the proposed approach relies less on the feedback of each individual service consumer, and more so on the \textit{collective wisdom} of all the service consumers that took part in transactions within a given MEC environment, as a whole. This helps the proposed approach train a trust bootstrapping model that generalizes better, producing more accurate results.

Furthermore, the specialized models derived from approaches such as Matrix Factorization are usually derived individually for a specific task requiring effort in modelling and design of a learning algorithm \cite{RN401}. This can hinder their generalizability. In contrast, the proposed approach can be applied to any arbitrary MEC-local trust bootstrapping strategy formulated as a convex optimization problem, by substituting it against $f_{S_{p}}^{q}$ in problem \eqref{eq:bt-predictor} and \eqref{eq:loss-function1}.

\subsubsection{Effectiveness of knowledge sharing}
The average prediction accuracy of the collaborative SVMs trained by the proposed approach and the non-collaborative SVMs trained by LO-TBM, MEC-local TT-SVD as well as MEC-local Wahab et al's showed 10.29\% and 21.7\% higher accuracy against the UNSW-NB15 and N-BaIoT datasets, respectively (see TABLE 3). The fact that both collaborative and non-collaborative SVMs were run under identical environmental settings, the above accuracy gain of the collaborative SVMs can be attributed to the effect of collaboration through knowledge sharing enforced by the proposed approach.

\subsubsection{Communication efficiency}
The results of our experiments on evaluating the communication efficiency revealed that the network stress imposed by the proposed approach is significantly less than that of the all cloud-based centralized SVM-based baseline models, which were trained under an identical environmental setting (see TABLE \ref{tab:comm-iters-and-time}).

\begin{table}[h!]
\centering
\caption{The number of rounds of communication between MEC and the centralized cloud layer observed at the point of achieving the maximum accuracy ($M=100$).}
% \bgroup
\def\arraystretch{1.2}
% \resizebox{0.45\textwidth}{!}{
\begin{tabular}{lcc}\toprule
    \multicolumn{1}{c}{\small \textbf{Model}} & \textbf{UNSW-NB15} & \textbf{N-BaIoT}\\\midrule
    GLB-TBM  & 159410 & 159410\\\midrule
    Global TT-SVD & 159410 & 159410\\\midrule
    Global Wahab's et al. & 159410 & 159410\\\midrule
    \textbf{Proposed approach}  & \textbf{2300} & \textbf{3800}\\\bottomrule
\end{tabular}
% }
\label{tab:comm-iters-and-time}
\end{table}

We attribute this difference in the number of communication iterations across the core networks of mobile networks to the localized nature in which the proposed approach tackles the training of its trust bootstrapping models. In other words, the proposed approach accumulates raw trust information within each MEC environment and crosses their network boundaries only for sharing knowledge among other MEC environments, via passing small messages (i.e. model parameters of the MEC-local trust bootstrapping models). In contrast, GLB-TBM and Global TT-SVD accumulates all its raw trust information within centralized cloud environments. This demands each record corresponding to transactions between IoT services and their consumers to be transmitted through to the centralized cloud-based data centers for training their respective trust bootstrapping strategies.

\subsubsection{Computational efficiency}

\begin{table}[h!]
\centering
\caption{Average time (in seconds) taken by the compared approaches to reach the underlying stopping criteria atop UNSW-NB15 and N-BaIoT datasets ($M=100$).}
% \bgroup
\def\arraystretch{1.2}
% \resizebox{.45\textwidth}{!}{
\begin{tabular}{lcc}\toprule
    \multicolumn{1}{c}{\textbf{Model}} & \textbf{UNSW-NB15} & \textbf{N-BaIoT}\\\midrule
    % \textbf{Model} & \textbf{UNSW-NB15} & \textbf{N-BaIoT}\\\midrule
    % LO-TBMs  & 85.32  & 70.1\\\midrule
    GLB-TBM & 4099.58 & 2612.12\\\midrule
    % MEC-local TT-SVD  & 76.32 & 62.52\\\midrule
    % MEC-local Wahab et al.'s & 50.21 & 45.93\\\midrule
    Global TT-SVD  & 505 & 496.51\\\midrule
    Global Wahab et al.'s & 142.77 & 121.34\\\midrule
    \textbf{Proposed approach}  & \textbf{251.14} & \textbf{490.32}\\\bottomrule
\end{tabular}
% }
% \captionsetup{justification=centering}

\label{tab:comp-eff}
\end{table}

The results of our experiments on evaluating the computational efficiency revealed that the proposed approach took the longest to reach its stopping criteria, when trained under an identical experimental setting as that of the other compared models (see TABLE \ref{tab:comp-eff}). There are multiple factors causing the above behaviour. Firstly, the proposed approach operates in three steps over multiple iterations as described in Section \ref{alg:solution}, each of which involves solving a distinct optimization problem. In contrast, all other approaches involve solving only one optimization problem to reach their respective optimality or stopping criteria. This, coupled with the differences of the underlying implementations of the solvers used by different approaches can be deemed to have caused the aforementioned difference in computational time. As it is evident in the existing literature, the computational time taken by ADMM-based approaches can be reduced by deriving closed-form solutions particularly on the \textit{z}-update involved in Algorithm 1 \cite{RN192}\cite{RN211}. Therefore, such measures can further be explored in order to improve the computational efficiency of the proposed approach.

\subsubsection{Scalability}

\begin{table}[h!]
\centering
\caption{Prediction accuracy (\%) of the evaluated distributed trust bootstrapping models atop UNSW-NB15 and N-BaIoT with the number of MEC environments in the MEC topology gradually increased.}
\begin{tabular}{cccccc} \toprule
    & & \multicolumn{4}{c}{No. of MEC environments} \\\cmidrule(lr){3-6}
    \textbf{Dataset} & \textbf{Model} & \textbf{100} & \textbf{150} & \textbf{200} & \textbf{250}\\\midrule
    UNSW-NB15 & LO-TBMs  & 84.87 & 84.81 & 84.71 & 84.19\\\cmidrule(lr){2-6}
    & GLB-TBM  & 87.28 & 87.05 & 87.1 & 86.99\\\cmidrule(lr){2-6}
    & MEC-local TT-SVD  & 79.59 & 82.85 & 82.38 & 82.07\\\cmidrule(lr){2-6}
    & MEC-local Wahab et al.'s & 76.32 & 79.5 & 77.27 & 70.87\\\cmidrule(lr){2-6}
    & \textbf{Proposed approach}  & \textbf{86.2} & \textbf{86.98} & \textbf{86.75} & \textbf{86.65}\\\midrule
    
    N-BaIoT & LO-TBMs & 82.16 & 83.29 & 82.05 & 81.87\\\cmidrule(lr){2-6}
    & GLB-TBM  & 83.18 & 83.08 & 82.9 & 82.75\\\cmidrule(lr){2-6}
    & MEC-local TT-SVD  & 79.15 & 80.7 & 80.74 & 80.21\\\cmidrule(lr){2-6}
    & MEC-local Wahab et al.'s & 62.52 & 63.2 & 65.03 & 64.21\\\cmidrule(lr){2-6}
    & \textbf{Proposed approach}  & \textbf{87.53} & \textbf{87.1} & \textbf{83.48} & \textbf{86.66}\\\bottomrule
\end{tabular}
\label{tab:avg-pred-acc-over-num-nodes}
\end{table}

% \begin{table}[h!]
% \centering
% \caption{Average prediction accuracy (\%) of the proposed approach atop UNSW-NB15 and N-BaIoT with the number of MEC environments in the MEC topology gradually increased.}
% % \resizebox{\linewidth}{!}{
% \begin{tabular}{ccccc} \toprule
%     & & \multicolumn{2}{c}{No. of MEC environments} \\\cmidrule(c){2-5}
%     Dataset & 100 & 150 & 200 & 250\\\midrule
%     UNSW-NB15 & 95.32 & 96.12 & 99.12 & 99.8\\\midrule
%     N-BaIoT & 78.6 & 84.34 & 85.11 & 82.4\\\bottomrule
% \end{tabular}
% % }
% \label{tab:avg-pred-acc-over-num-nodes}
% \end{table}

\begin{table}[h!]
\centering
\caption{Prediction accuracy (\%) of the evaluated distributed trust bootstrapping models atop UNSW-NB15 and N-BaIoT with volumes of data in each MEC environment gradually increased.}

\begin{tabular}{cccccc} \toprule
    & & \multicolumn{4}{c}{No. of MEC environments} \\\cmidrule(lr){3-6}
    \textbf{Dataset} & \textbf{Model} & \textbf{25\%} & \textbf{50\%} & \textbf{75\%} & \textbf{100\%}\\\midrule
    UNSW-NB15 & LO-TBMs & 85.02 & 85.28 & 85.62 & 85.96\\\cmidrule(lr){2-6}
     & Global-TBM  & 87.32 & 87.35 & 87.33 & 87.29\\\cmidrule(lr){2-6}
    & MEC-local TT-SVD & 82.40 & 80.11 & 81.46 & 83.41\\\cmidrule(lr){2-6}
    & MEC-local Wahab et al. & 78.12 & 79.50 & 77.27 & 70.87\\\cmidrule(lr){2-6}
    & \textbf{Proposed approach} & \textbf{86.02} & \textbf{86.05} & \textbf{86.12} & \textbf{86.30}\\\midrule
    
    N-BaIoT & LO-TBMs & 82.05 & 81.94 & 82.40 & 82.19\\\cmidrule(lr){2-6}
         & Global-TBM & 83.07 & 83.06 & 83.06 & 83.10\\\cmidrule(lr){2-6}
    & MEC-local TT-SVD & 78.49 & 80.19 & 75.83 & 75.50\\\cmidrule(lr){2-6}
    & MEC-local Wahab et al. & 63.70 & 64.12 & 67.11 & 68.70\\\cmidrule(lr){2-6}
    & \textbf{Proposed approach} & \textbf{82.69} & \textbf{82.83} & \textbf{82.90} & \textbf{82.99}\\\bottomrule
\end{tabular}

\label{tab:avg-pred-acc-over-num-nodes}
\end{table}

% \begin{table}[h!]
% \centering
% \caption{Computational time (in seconds) taken by the evaluated distributed trust bootstrapping models atop UNSW-NB15 and N-BaIoT with volumes of data in each MEC environment was gradually increased.}
% \resizebox{\linewidth}{!}{
% \begin{tabular}{cccccc} \toprule
%     & & \multicolumn{4}{c}{No. of MEC environments} \\\cmidrule(c){3-6}
%     \textbf{Dataset} & \textbf{Model} & \textbf{25\%} & \textbf{50\%} & \textbf{75\%} & \textbf{100\%}\\\midrule
%     UNSW-NB15 & LO-TBMs & 4.31 & 5.24 & 6.12 & 7.19\\\cmidrule{2-6}
%      & Global-TBM & 11110.13 & 20495.71 & 26586.91 & 33217.3\\\cmidrule{2-6}
%     & MEC-local TT-SVD & 18.2 & 24.79 & 31.32 & 38.33\\\cmidrule{2-6}
%     & MEC-local Wahab et al.'s & - & - & - & -\\\cmidrule{2-6}
%     & \textbf{Proposed approach}  & \textbf{-} & \textbf{-} & \textbf{-} & \textbf{-}\\\midrule
    
%     N-BaIoT & LO-TBMs & 5.72 & 7.29 & 9.42 & 11.78\\\cmidrule{2-6}
%          & Global-TBM & 4434.15 & 9025.64 & 11531.56 & 17063.24\\\cmidrule{2-6}
%     & MEC-local TT-SVD & 17.37 & 22.32 & 28.87 & 35.75\\\cmidrule{2-6}
%     & MEC-local Wahab et al.'s & - & - & - & -\\\cmidrule{2-6}
%     & \textbf{Proposed approach} & \textbf{-} & \textbf{-} & \textbf{-} & \textbf{-}\\\bottomrule
% \end{tabular}
% }
% \label{tab:avg-pred-acc-over-num-nodes}
% \end{table}

\begin{table}[h!]
\centering
\caption{Average communication iterations taken by the proposed approach atop UNSW-NB15 and N-BaIoT when the number of MEC environments in the topology gradually increased.}
\begin{tabular}{ccccc}
\toprule
\textbf{Dataset} & \textbf{100} & \textbf{150} & \textbf{200} & \textbf{250} \\
\midrule
UNSW-NB15 & 2300 & 5550 & 6800 & 7500 \\
N-BaIoT   & 3800 & 5550 & 5400 & 9750 \\
\bottomrule
\end{tabular}
\label{tab:comm-iters-over-num-nodes}
\end{table}

\begin{figure}[t]
\centering
\includegraphics[width=0.6\textwidth]{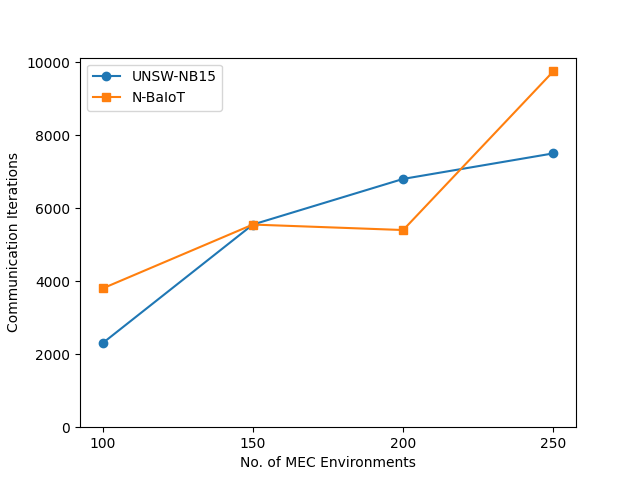}
\caption{Change in the number of communication iterations required to achieve the maximum accuracy when the number of MEC environments in the underlying MEC topology was gradually increased.} \label{fig1}
\end{figure}

The results of our experiments showed that the proposed approach was comparatively more scalable on the number of communication iterations taken till convergence, atop both datasets (see TABLE \ref{tab:comm-iters-over-num-nodes}). In other words, it was observed that When the number of MEC environments in the underlying simulated MEC topology was increased in batches of 50, the number of communication iterations of the proposed approach grew sub-linearly. Such a behaviour can often be deemed desirable in order to avoid excessive network stress on the core networks of the mobile network providers in the presence of growing MEC environments.

\section{Conclusion and Future Work}\label{sec:conclusion}

This paper proposes a data-driven context-aware strategy to bootstrap trustworthiness of lesser-known Mobile Edge Computing (MEC)-based IoT services. In addition, this work also aims to tackle the \textit{data sparsity} related problems arising in the aforementioned context due to the  split-nature in which trust information is gathered across distributed MEC environments hindering the ability to reliably bootstrap trustworthiness of \textit{lesser-known} IoT services. To address these challenges, we first formally model the problem of trust bootstrapping in the aforementioned setting. We 
% \sout{then use the Alternating Direction Method of Multipliers (ADMM) to come up with a distributed machine learning based approach} 
then introduced a distributed solution for trust bootstrapping in MEC-based IoT services based on the Alternating Direction Method of Multipliers that also allows knowledge sharing among similar trust regions to counter the effects of data sparsity. 
%aligned with the goals of MEC paradigm. %In addition, the proposed approach also enables knowledge sharing among similarly-poised MEC environments that carry similar \textit{trust regions}, which are identified and linked together automatically. 
The feasibility of our approach was affirmed via simulated experiments conducted atop 
% carefully 
curated data extracted from two popular IoT datasets. 

Our future work aims to build on the proposed work to come up with a more holistic approach to bootstrap trustworthiness of IoT services in the considered problem setting. This includes investigating the problem of bootstrapping the trustworthiness of IoT services when there is no sufficient trust information or well-known services available, as well as, evaluating how the proposed approach behaves in more volatile environments where the trust information could not be fully trusted. In addition, we also hope to investigate the approaches to better equip the proposed approach in terms of handling the dynamicity associated with MEC environments with respect to user mobility, service availability, etc.

% Can use something like this to put references on a page
% by themselves when using endfloat and the captionsoff option.
% \ifCLASSOPTIONcaptionsoff
%   \newpage
% \fi

% \vspace{-4mm}
\bibliographystyle{plain}
\bibliography{tii-2023}

\end{document}